\documentclass[journal]{IEEEtran}

\usepackage{graphicx}
\usepackage{cite}
\usepackage{soul}
\usepackage{framed}
\usepackage{epsfig,mathptmx,times,amsmath,amssymb,color,flushend,gensymb,subfigure,fancyhdr,subfigure}
\usepackage{framed}
\definecolor{shadecolor}{rgb}{1,0.8,0.3}

\ifCLASSINFOpdf
\else
\fi

\begin{document}

\title{Robust Tube-based Decentralized Nonlinear Model Predictive Control of an Autonomous Tractor-Trailer System}

\author{Erkan~Kayacan,~\IEEEmembership{Student Member, IEEE,} ~~Erdal~Kayacan,~\IEEEmembership{Senior Member, IEEE,} ~~Herman~Ramon~\IEEEmembership{}~and~Wouter~Saeys~\IEEEmembership{}% <-this % stops a space
\thanks{Erkan Kayacan, Herman Ramon and Wouter Saeys are with the Division of Mechatronics, Biostatistics and Sensors, Department of Biosystems, University of Leuven (KU Leuven), Kasteelpark Arenberg 30, B-3001 Leuven, Belgium.
e-mail: {\tt\small \{erkan.kayacan, herman.ramon, wouter.saeys\}@biw.kuleuven.be }}
\thanks{Erdal Kayacan is with the School of Mechanical and Aerospace Engineering, Nanyang Technological University, 639798, Singapore.
e-mail: {\tt\small erdal@ntu.edu.sg }}
}

\markboth{\textbf{PREPRINT VERSION:} IEEE/ASME TRANSACTIONS ON MECHATRONICS, vol. 23, no. 1, pp. 197-205, Jan. 2015.}
{Shell \MakeLowercase{\textit{et al.}}: Bare Demo of IEEEtran.cls for Journals}
\maketitle

\begin{abstract}
This paper addresses the trajectory tracking problem of an autonomous tractor-trailer system by using a decentralized control approach. A fully decentralized model predictive controller is designed in which interactions between subsystems are neglected and assumed to be perturbations to each other. In order to have a robust design, a tube-based approach is proposed to handle the differences between the nominal model and real system. Nonlinear moving horizon estimation is used for the state and parameter estimation after each new measurement, and the estimated values are fed the to robust tube-based decentralized nonlinear model predictive controller. The proposed control scheme is capable of driving the tractor-trailer system to any desired trajectory ensuring high control accuracy and robustness against neglected subsystem interactions and environmental disturbances. The experimental results show an accurate trajectory tracking performance on a bumpy grass field.
\end{abstract}


\IEEEpeerreviewmaketitle

\section{Introduction}

\IEEEPARstart{A}{n} autonomous tractor with a trailer attached to it is a complex mechatronic system in which the overall system dynamics can be divided into, at least, three subsystems: the longitudinal dynamics, the yaw dynamics of the tractor and the yaw dynamics of the trailer. Moreover, there exist interactions between these subsystems. First, since the tractor and the trailer are mechanically coupled to each other, a steering angle input applied to the tractor affects not only the yaw dynamics of the tractor but also the yaw dynamics of the trailer. Second, the same hydraulic oil is used in the overall system which makes that an input to one of the three subsystems also affects the others. Finally, the diesel engine rpm has a direct effect on the hydraulic oil flow. This implies that a manipulation on the diesel engine rpm affects all the subsystem dynamics.

Various implementation examples to control tractor with/without trailer system are seen in literature. In order to follow straight lines, model reference adaptive control was proposed for the control of a tractor configured with different trailers in \cite{Derrick}, and a linear quadratic regulator was used to control a tractor-trailer system in \cite{karkeejournal}. Both controllers have been designed based on dynamic models. However, since these dynamic models are derived with a small steering angle assumption, they are not suitable for curvilinear trajectory tracking. For curvilinear trajectories, NMPC was proposed for the control of a tractor-trailer system in \cite{Backman2012}. Extended Kalman filter (EKF) was used to estimate the yaw angles of the tractor and trailer. However, the effects of side-slip were neglected. In \cite{TomKraus}, the states and parameters of a tractor including the wheel slip and side-slip were estimated with nonlinear moving horizon estimation (NMHE) and fed to a nonlinear MPC. As a model-free approach, a type-2 fuzzy neural network with a sliding mode control theory-based learning algorithm was proposed to control of a tractor in \cite{erdalt2fnn}.

The aforementioned interactions make the control of complex mechatronic systems challenging. One candidate solution is the use of a centralized control approach, \emph{e.g.} centralized model predictive control (CeMPC). However, the main disadvantage of the centralized control approach is that the centralized control of such systems using a plant-wide model may not be computationally feasible since the optimization process of a multi-input-multi-output system is a time consuming task \cite{Liu2010,Lynch}. As a simpler alternative solution, decentralized MPC (DeMPC) can be preferred in which the global optimization problem is divided into smaller pieces resulting in simpler and tractable optimization problems. In this method, local control inputs are computed using only local measurements, and it reduces the order of the models to that specific local subsystem \cite{Scattolini2009}. The main drawback of this approach is that it neglects the system interactions and has to deal with them as if they are disturbances. If the subsystem interactions are not very strong in a complex mechatronic system, this approach can be preferred.

De(N)MPC has recently been studied by several researchers as it requires simpler optimization problems when compared to its centralized counterpart. In \cite{Magni}, a fully decentralized structure has been studied in which the overall system is nonlinear, discrete time and no information can be exchanged between local controllers. Whereas the system is also discrete-time and nonlinear in \cite{Raimondo}, each subsystem is locally controlled with an MPC algorithm guaranteeing the input-to-state stability property. Unlike \cite{Magni} and \cite{Raimondo}, there is a partial exchange of information between subsystems in \cite{Dunbar,erkanDiNMPC}. It is to be noted that the most real world implementations are similar to the case in \cite{Magni} and \cite{Raimondo} in which the systems are fully decentralized. In this paper, we also focus on such a design that there is no information exchange between the subsystems.

Although (N)MPC has caught noticeable attention from researchers for its ability to handle constraints as well as nonlinearities in multi-input-multi-output systems, robust stability can only be obtained if the nominal system is inherently robust and the state estimation errors are sufficiently small \cite{Mayne}. Unfortunately, predictive controllers are not always inherently robust \cite{Rawlings}. One approach to deal with this drawback is the use of robust (N)MPC design methods, \emph{e.g.} by taking the state estimation error into account. In \cite{Mayne2004}, a tube-based MPC has been proposed which generates the inputs to the system based-on the measurements coming from the nominal model. The aforementioned structure was criticised because the method does not take the outputs of the real-time system into account. As an alternative approach, a novel tube-based MPC, which tries to minimize the cost function with respect to the outputs of the real-time system, has been proposed for state feedback in \cite{Mayne2005} and output feedback in \cite{Mayne2006}. In earlier studies, the tube-based approach was formulated only for the discrete-time and linear MPC case. Recently, it has been extended to the continuous-time case in \cite{Farina2012} and the nonlinear case in \cite{Maynenonlineartube,Yu2013}. In another study, tube-based MPC was proposed for the control of large-scale systems with a distributed control scheme in which a decentralized static state-feedback controller is used for the control of each subsystem \cite{Riverso2012}.  In this paper, the approach in \cite{Riverso2012} has been extended to the nonlinear and decentralized MPC case.

There exist successful real-time implementations of tube-based MPC in literature: A nonlinear model can be linearized around a working point and described as a linear system with additive disturbances. In \cite{Limon2010}, the robustness of the tube-based MPC has been elaborated against significantly changing working points based-on a single optimization problem. The experimental results on a quadruple-tank plant show the stability and offset-free tracking of the control algorithm. The real-time examples on tube-based approach have been extended to motion planning and trajectory tracking of mobile robots in \cite{Farrokhsiar2013,Gonzalez2011}. Whereas the ancillary control law was linear-time-invariant in \cite{Gonzalez2011}, the approach has been extended by using an adaptive state feedback gain in \cite{Gonzalez2011adaptive} for the trajectory tracking problem of mobile robots.

\emph{Contribution of this paper: }
In this study, a fast, robust, tube-based decentralized NMPC has been implemented and tested in real-time with respect to its potential to obtain fast, accurate and efficient trajectory tracking of a tractor-trailer system. To succeed, the following selections have been made:
\begin{itemize}
  \item The use of C++ source files to realize the control algorithm in real-time,
  \item The use of the decentralized control algorithm instead of a centralized one,
  \item A simple solution for the optimization problems in NMPC and NMHE is used in which the number of Gauss-Newton iterations is limited to 1.
  \item A practical mechatronic system, illustrating how control, sensing and actuation can be integrated to achieve an intelligent system, is designed and presented.
\end{itemize}

This paper is organized as follows: The experimental set-up and the kinematic tricycle model of the system are presented in Section \ref{System}. The basics of the implemented robust tube-based DeNMPC  approach and the learning process by using NMHE are described in Section \ref{denmpcnmheN}.  The experimental results are presented in Section \ref{realtime}. Finally, some conclusions are drawn from this study in Section \ref{Conc}.

%\hfill mds
%\hfill December 27, 2012
%%%%%%%%%%%%%%%%%%%%%%%%%%%%%%%%%%%%%%%%%%%%%%%%%%%%%%%%%%%%%%%%%%%%%%%%%%%%%%%%%%%%%%%%%%%%%%%%%%%%%%%%%%%%%%%%%%%%%%%%%%%%%%%%%%%%%%%%%%%%%%%%%%%%%%%%%%%%%%%%

\section{Autonomous Tractor-trailer System}\label{System}
\subsection{Experimental Set-up Description}

\begin{figure}[h!]
\centering
  \includegraphics[width=3in]{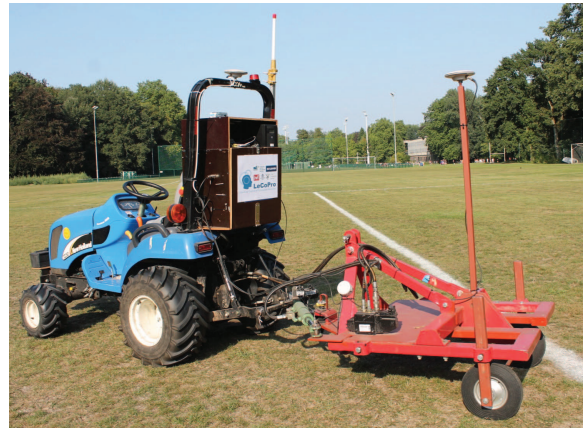}\\
  \caption{The tractor-trailer system}
  \label{tractor1}
\end{figure}

The global aim of the real-time experiments in this paper is to track a space-based trajectory with the small agricultural tractor-trailer system shown in Fig. \ref{tractor1}. Two GPS antennas are located straight up the center of the tractor rear axle and the center of the trailer to provide highly accurate positional information. They are connected to a Septentrio AsteRx2eH RTK-DGPS receiver (Septentrio Satellite Navigation NV, Belgium) with a specified position accuracy of 2 cm at a 5-Hz sampling frequency. The Flepos network supplies the RTK correction signals via internet by using a \emph{Digi Connect WAN 3G} modem.

The GPS receiver and the internet modem are connected to a real time operating system (PXI platform, National Instrument Corporation, USA) through an RS232 serial communication. The PXI system acquires the steering angles and the GPS data, and controls the tractor-trailer system by applying voltages to the actuators. A laptop connected to the PXI system by WiFi functions as the user interface of the autonomous tractor. The control algorithms are implemented in $LabVIEW^{TM}$ version 2011, National Instrument, USA. They are executed in real time on the PXI and updated at a rate of 5-Hz.

The robust tube-based DeNMPC  calculates the desired steering angles for the front wheels of the tractor and the trailer, respectively. These reference signals are then sent to two low level controllers, PI controllers in our case, which provide the low level control of the steering mechanisms. While the position of the front wheels of the tractor is measured using a potentiometer mounted on the front axle yielding a position measurement resolution of $1$ degree, the position of the electro-hydraulic valve on the trailer is measured by using an inductive sensor with $1$ degree precision.

\begin{figure}[h!]
\centering
\subfigure[ ]{
\includegraphics[width=1.5in]{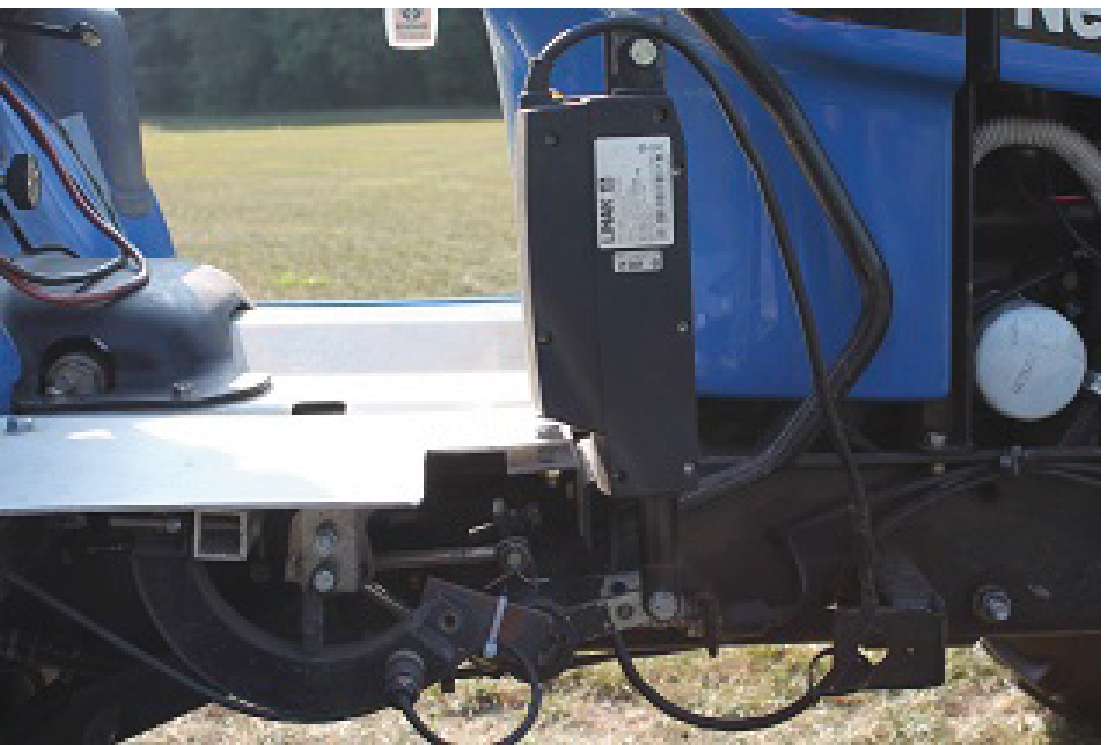}
\label{hydrostate_EMV}
}
\subfigure[ ]{
\includegraphics[width=1.5in]{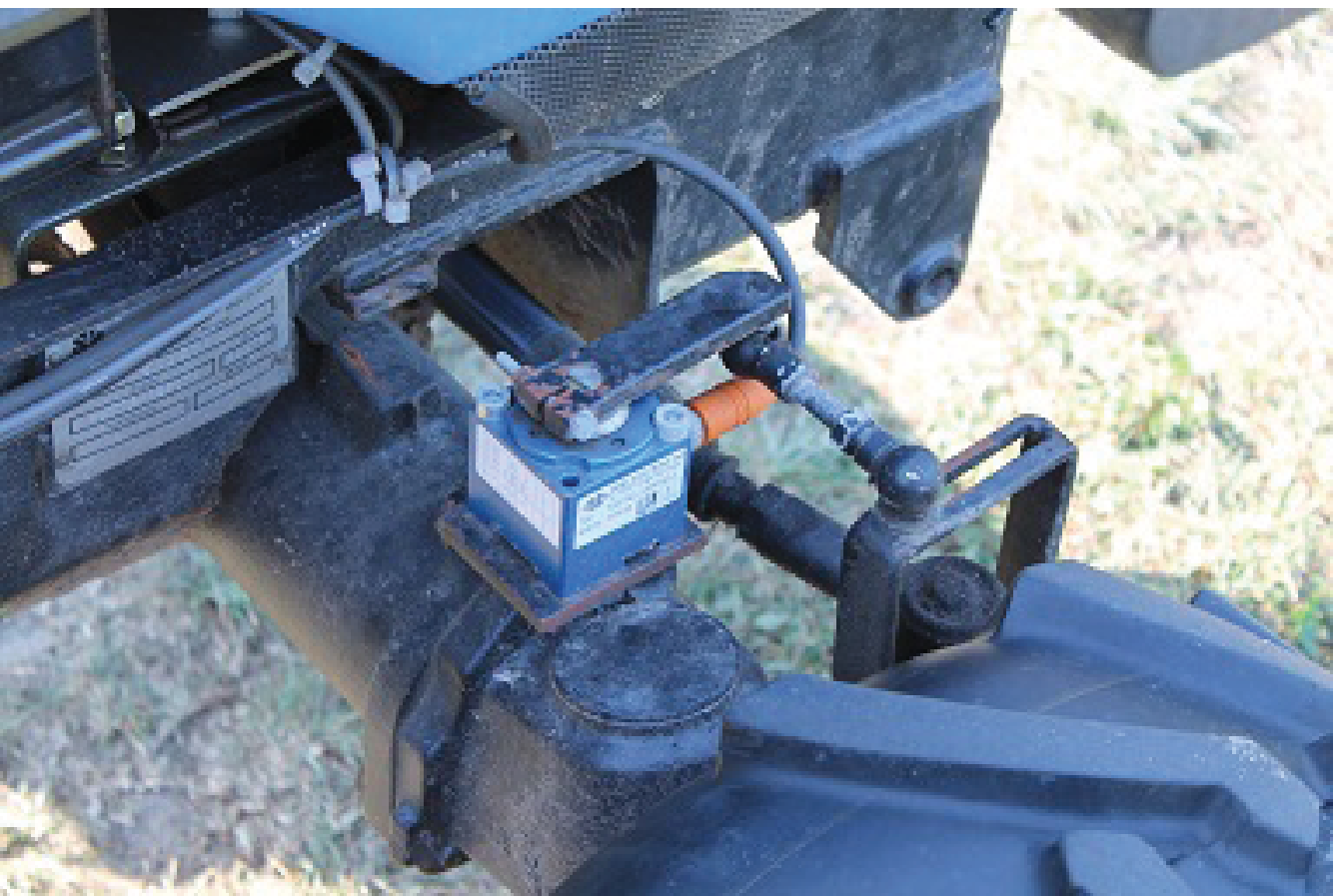}
\label{steering-potantiometer}
}
\subfigure[ ]{
\includegraphics[width=2in]{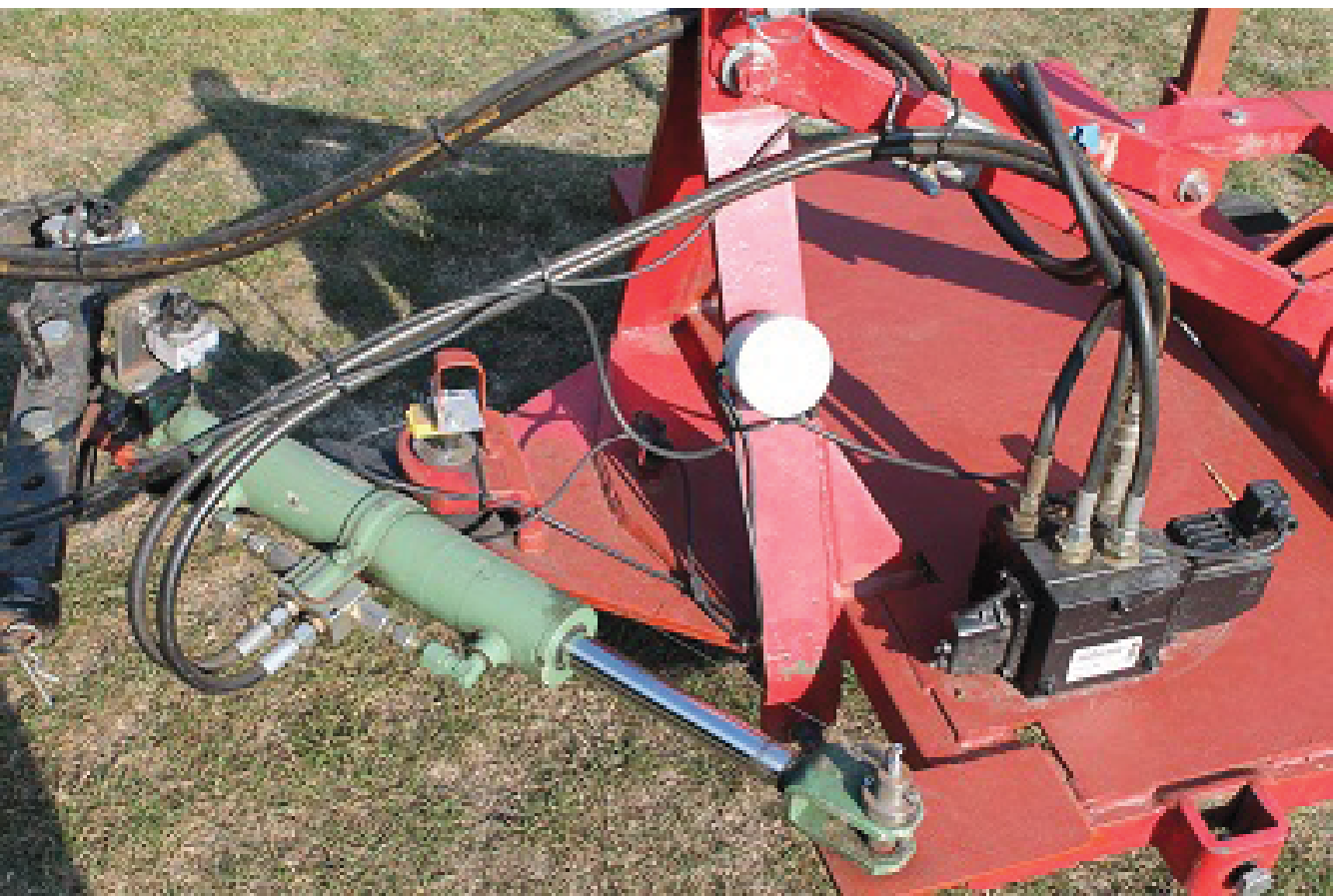}
\label{Trailer actuator}
}
\caption[Optional caption for list of figures]{(a) Hydrostat electro-mechanical valve (b) Steering angle potentiometer (c) Trailer actuator}
\label{sensors}
\end{figure}

The speed of the tractor is controlled through an electromechanical actuator connected to the hydrostat pedal connected to the variable hydromotor. The wheel speed is controlled by a cascade system with two PID controllers, where the inner loop controls the hydrostat pedal position to the reference position requested by the outer loop. Figure \ref{sensors} shows the hydrostat electro-mechanical valve (Fig. \ref{hydrostate_EMV}), the steering angle potentiometer (Fig. \ref{steering-potantiometer}) and the trailer actuator (Fig. \ref{Trailer actuator}), respectively.

\subsection{Kinematic Tricycle Model}

\begin{figure}[h!]
\centering
  \includegraphics[width=2.5in]{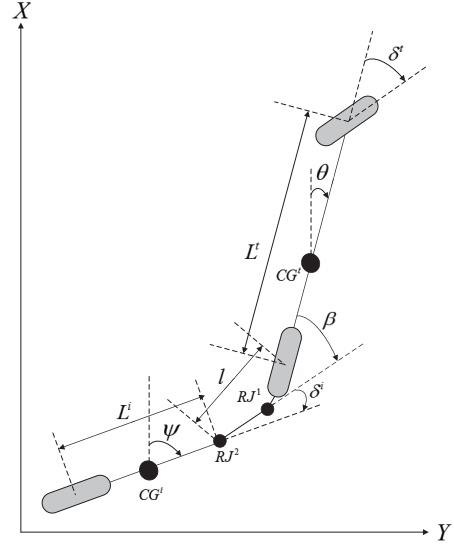}\\
  \caption{Schematic illustration of tricycle model for an autonomous tractor-trailer system}
  \label{kinematic}
\end{figure}

The model for the autonomous tractor-trailer system is \emph{an adaptive kinematic model} neglecting the dynamic force balances in the equations of motion. The model used here is an extension of the ones used in \cite{karkeejournal,karkeephd}. The extensions are the additional three slip parameters ($\mu$, $\kappa$ and $\eta$) and the definition of the yaw angle difference between the tractor and the trailer by using two angle measurements ($\alpha$ and $\beta$) instead of one angle measurement. A dynamic model would, of course, represent the system behaviour with a better accuracy, but the investment for building such a model through multibody modelling and system identification would be considerably higher \cite{erkan2013acc,erkan2014modellemejournal}. Moreover, a dynamic model would increase the computational burden in the optimization process in DeNMPC. The schematic diagram of the autonomous tractor-trailer system is presented in Fig. \ref{kinematic}.

The equations of motion of the system to be controlled are as follows:
\small
\begin{eqnarray}\label{kinematicmodel}
\left[
  \begin{array}{c}
  \dot{x}^{t}  \\
  \dot{y}^{t}  \\
  \dot{\theta} \\
  \dot{x}^{i}  \\
  \dot{y}^{i}  \\
  \dot{\psi}   \\
  \end{array}
  \right]
 =
\left[
  \begin{array}{c}
  \mu v \cos{(\theta)} \\
 \mu v \sin{(\theta)}  \\
\frac{\mu v \tan{ (\kappa \delta^{t}) } }{L^t} \\
 \mu v \cos{(\psi)}  \\
 \mu v \sin{(\psi)}  \\
\frac{\mu v }{L^i}\big(\sin{(\eta \delta^{i} + \beta)} - \frac{l }{L^t} \tan{ (\kappa \delta^{t})} \cos{(\eta \delta^{i} + \beta)} \big)  \\
  \end{array}
  \right]
\end{eqnarray}
\normalsize
where ${x}^{t}$ and $y^{t}$ represent the position of the tractor, $\theta$ is the yaw angle of the tractor, ${x}^{i}$ and $y^{i}$ represent the position of the trailer, $\psi$ is the yaw angle of the trailer, $v$ is the longitudinal speed of the system. Since the tractor and trailer rigid bodies are linked by two revolute joints at a hitch point, the tractor and the trailer longitudinal velocities are coupled to each other.  The steering angle of the front wheel of the tractor is represented by $\delta^{t}$, $\beta$ is the hitch point angle between the tractor and the drawbar at $RJ^{1}$; $\delta^{i}$ is the steering angle between the trailer and the drawbar at $RJ^{2}$; $\mu$, $\kappa$ and $\eta$ are the slip coefficients for the wheel slip of the tractor, side-slip for the tractor and side-slip for the trailer, respectively. It is to be noted that the slip parameters can only get values between zero and one. While a wheel slip of one indicates that the wheel and tractor velocities are the same, a ratio of zero indicates that the wheels are skidding on the surface, i.e., the tractor is no longer steerable \cite{Kayacan2009,Topalov2011}.

The physical parameters that can be directly measured are as follows: The distance between the front axle of the tractor and the rear axle of the tractor $L^t (1.4 m)$, the distance between $RJ^2$ and the rear axle of the trailer $L^i (1.3 m)$ and the distance between the rear axle of the tractor and $RJ^2$ $l(1.1 m)$, respectively.

%%%%%%%%%%%%%%%%%%%%%%%%%%%%%%%%%%%%%%%%%%%%%%%%%%%%%%%%%%%%%%%%%%%%%%%%%%%%%%%%%%%%%%%%%%%%%%%%%%%%%%%%%%%%%%%%%%%%%%%%%%%%%%%%%%%%%%%%%%%%%%%%%%%%%%%%%%%%%%%%
\section{Nonlinear Moving Horizon Estimation and Decentralized Nonlinear Model Predictive Control} \label{denmpcnmheN}

\subsection{Nonlinear Moving Horizon Estimation}
As any type of (N)MPC requires information on the system states, these have to be either directly measured or estimated. In practical applications, it is typically impossible to measure all states directly. Therefore, it is generally necessary to estimate some states or unknown model parameters online when working with (N)MPC. The most commonly used method for state and parameter estimation is the EKF. However, the main disadvantage of the EKF approach is that this method cannot deal with the constraints on the states or parameters (e.g. no negative wheel slip). To overcome this limitation of the EKF moving horizon estimation (MHE) has been proposed as an optimization based state estimator \cite{TomKraus}. In this paper, an alternative method, NMHE has been preferred since it treats the state and the parameter estimation within the same problem and also constraints can be incorporated. The constraints play an important role in the autonomous tractor-trailer system. For instance, the slip coefficients cannot be larger than $1$.

The NMHE problem can be formulated as follows:
\begin{equation}
 \begin{aligned}
 & \underset{x(.),p,u(.)}{\text{min}}
 & & \int^{t_k}_{t_k-t_h} \| y_m (t) - h \big(x(t),u(t),p \big) \|^{2}_{H} dt  \\ &
 && + \left\|
  \begin{array}{c}
    \hat{x} (t_k-t_h) - x (t_k-t_h)  \\
    \hat{p} - p
 \end{array}
 \right\| ^{2}_{P} \\
 & \text{subject to}
 &&  \dot{x}(t) = f \big(x(t),u(t),p \big) \\
 &&& x_{min} \leq x(t) \leq x_{max} \\
 &&& p_{min} \leq p \leq p_{max} \;\; \text{for all} \;\; t \in [t_k - t_h ,t_k]\\
  \end{aligned}
    \label{mhe}
\end{equation}
where $y_m$ and $h$ are the measured output and measurement function, respectively. Deviations of the first states in the moving horizon window and the parameters from priori estimates $\hat{x}$ and $\hat{p}$ are penalised by a symmetric positive definite matrix $P$. Moreover,  deviations of the predicted system outputs and the measured outputs are penalised by symmetric positive definite matrix $H$ \cite{Ferreau}. Upper and lower bounds on the model parameters are represented by parameters $p_{min}$ and $p_{max}$, respectively.

The last term in the objective function in \eqref{mhe} is called the arrival cost. The reference estimated values $\hat{x} (t_k-t_h)$ and $\hat{p}$ are taken from the solution of NMHE at the previous estimation instant. In this paper, the matrix $P$ for the arrival cost has been chosen as a so-called smoothed EKF-update based on sensitivity information obtained while solving the previous NMHE problem \cite{Robertson}. The contributions of the past measurements to the covariance matrix $P$ are downweighted by a process noise covariance matrix $D_{update}$ which must be available. The calculation of $P$ can be found in \cite{Tomphdthesis,Robertson}.

\subsection{Decentralized Nonlinear Model Predictive Control}\label{sectionDeNMPC}

In a single-input-single-output control scheme, the aim is to follow a constant or time-varying reference by using one control variable. However, in a multi-input-multi-output control, multiple interacting states are controlled by using multiple control variables. This makes it considerably more challenging to design an appropriate control scheme for such systems. When a process model is available, all the interactions between the different subsystems can be taken into account by using a model-predictive control approach.  However, as many of these MIMO systems, such as the tractor-trailer system investigated in this study are nonlinear in nature, these cannot be conveniently controlled with linear MPC. This results in a necessity of the combination of a nonlinear model and an MPC which is referred to as NMPC.

In order to be able to design a DeNMPC, a partitioned model of full system should be available derived from partitioning methods as non-overlapping decomposition or completely overlapping decomposition. However, considering the kinematic model in \eqref{kinematicmodel}, the first three states of \eqref{kinematicmodel} are the state equations of the tractor while the last three states are the state equations of the trailer. Thus, the equations of motion for the tractor-trailer system represented in \eqref{kinematicmodel} are naturally decoupled so that partitioning methods are not needed for our system. Even if there exist several interactions in the real-time application, the only subsystem interaction in \eqref{kinematicmodel} is that the steering angle of the tractor has influence on the yaw angle of the trailer. Since the subsystem model has to consist of only its states and inputs in DeNMPC, the effect of the steering angle of the tractor on the yaw angle of the trailer will be neglected. As a result, the new equation for the yaw angle of the trailer is written as follows:
\begin{equation}\label{kinematicmodelDe}
  \dot{\psi} = \frac{\mu v }{L^i}\big(\sin{(\eta \delta^{i} + \beta)} \big)
\end{equation}

%In NMPC, it is to be noted that the computational load in the optimization process increases when the number of the states of the system increases. For this reason, one of the advantages of the DeNMPC is to lower the computational burden by decoupling the subsystems. Once the subsystems are decoupled from each other, the DeNMPC algorithm will not need a scalable memory load. A second advantage is that less communication between subsystems results in the reduction of the transmission delays and overloads.

For the formulation of DeNMPC, it is assumed that the plant comprises N subsystems to give the general formulation for DeNMPC.

\subsubsection{System Model}
A nonlinear system model consisting of N subsystems is written for each subsystem as follows:
\begin{equation}
\dot{x}_{i}(t) = f_{i} (x_{i}(t),u_{i}(t)) + g_{i} (x(t),u(t)) + d_{i} (t), \;\;\;  i \in \mathbb{I}_{1:N}
\label{nonlinearmodel}
\end{equation}
where $x_i$ $\in$ $\mathbb{R}^{n_{i}}$, $u_{i}$ $\in$ $\mathbb{R}^{m_{i}}$, and $d_i$ $\in$ $\mathbb{R}^{n_{i}}$ are respectively the state, the input and the disturbance of the $i^{th}$ subsystem. The influence of the $i^{th}$ subsystem and the influence of the other subsystems on the $i^{th}$ subsystem are described by $f_{i}$ and $g_{i}$ functions that are continuously differentiable, respectively.

At each time-step, the states and the inputs have to satisfy:
\begin{equation}
x_{i} \in \mathbb{X}_{i}, \;\;\; u_{i} \in \mathbb{U}_{i}
\end{equation}
where $\mathbb{X}_{i} \subseteq  \mathbb{R}^{n_{i}}$ is closed, $\mathbb{U}_{i} \subseteq  \mathbb{R}^{m_{i}}$ is compact and each set contains the origin in its interior point. The constraints for each input are defined uncoupled because the feasible regions of the inputs do not affect each other. The disturbance $d_{i}$ is assumed to be bounded,
\begin{equation}
d_{i} \in \mathbb{D}_{i}
\end{equation}
where $\mathbb{D}_{i} \subseteq  \mathbb{R}^{n_{i}}$ is compact and contains the origin in its interior point.

From \eqref{nonlinearmodel}, the nominal system for each subsystem is obtained by neglecting the subsystem interaction $g_{i} (x(t),u(t))$ and the disturbance $d_{i} (t)$ as follows:
\begin{equation}
\dot{\bar{x}}_{i}(t) = f_{i} (\bar{x}_{i}(t), \bar{u}_{i}(t)), \;\;\; i \in \mathbb{I}_{1:N}
\label{nominalmodel}
\end{equation}
where $\bar{x}_{i}$ $\in$ $\mathbb{R}^{n_{i}}$ and $\bar{u}_{i}$ $\in$ $\mathbb{R}^{m_{i}}$ are the nominal state and input, respectively.

\subsubsection{Objective Functions}
The stage cost and the terminal penalty are respectively written for each subsystem $i \in \mathbb{I}_{1:N}$ as follows:
\begin{eqnarray}
V_{iSC} (\bar{x}_i, \bar{u}_i) & = & \| \bar{x}_{ir} (t) - \bar{x}_i (t) \|^{2}_{Q_i} + \| \bar{u}_{ir} (t) - \bar{u}_i (t) \|^{2}_{R_i}  \\
V_{iTP} (\bar{x}_i)      & = & \| \bar{x}_{ir} (t_k+t_h) - \bar{x}_i (t_k+t_h) \|^{2}_{S_i}
\end{eqnarray}
where $Q_i \in \mathbb{R}^{n_i \times n_i}$, $R_i \in \mathbb{R}^{m_i \times m_i}$ and $S_i \in \mathbb{R}^{n_i \times n_i}$ are weighting matrices being symmetric and positive definite, $\bar{x}_{ir}$ and $\bar{u}_{ir}$ are the references for the states and the inputs, $\bar{x}_i$ and $\bar{u}_i$ are the states and the inputs, $t_k$ stands for the current time, $t_h$ is the prediction horizon.

The objective function for each subsystem $i \in \mathbb{I}_{1:N}$ is written as follows:
\begin{equation}
 \begin{aligned}
 & V_i (\bar{x}_i, \bar{u}_i)=
 & & \int^{t_k+t_h}_{t_k} \big(V_{iSC} (\bar{x}_i, \bar{u}_i)\big) dt + V_{iTP} (\bar{x}_i)  \\
 & && \forall t \in [t_k, t_k+t_h] \\
  \end{aligned}
\end{equation}

\subsubsection{Formulation of DeNMPC}
The plant objective function is written as follows:
\begin{equation}
 \begin{aligned}
 & \underset{\bar{x}_i(.), \bar{u}_i(.)}{\text{min}}
 & & V_{i}(\bar{x}_i,\bar{u}_i) \\
 & \text{subject to}
 && \bar{x}_i(t_k) = \hat{x}_i(t_k) \\
 && & \dot{\bar{x}}_i(t) = f_i \big(\bar{x}_i(t), \bar{u}_i(t) \big) \\
 && & \bar{x}_{i_{min}} \leq \bar{x}_i(t) \leq \bar{x}_{i_{max}} \\
 && & \bar{u}_{i_{min}} \leq \bar{u}_i(t) \leq \bar{u}_{i_{max}} \;\;\; \forall t \in [t_k, t_k+t_h] \\
  \end{aligned}
  \label{DeNMPC}
\end{equation}
where  $V_{i}$ is the plant objective function. Moreover, upper and lower bounds on the state and the input are represented by $\bar{x}_{i_{min}}$, $\bar{x}_{i_{max}}$, $\bar{u}_{i_{min}}$ and $\bar{u}_{i_{max}}$. The stability proof of DeNMPC can be found in \cite{Magni,Balas,Raimondo}.

\subsection{Robust Tube-based Decentralized Nonlinear Model Predictive Control}\label{rDeNMPC}

As can be seen from \eqref{nonlinearmodel}, the nonlinear model for each subsystem consists of its state, its input, the influence of other subsystems and the disturbance. However, the nominal model in \eqref{nominalmodel} does not consist of the subsystem interactions. In the decentralized control approach the effects of interconnections are treated as perturbations. For this reason, the uncertainty between the nominal model and the real system can result in poor performance for real-time applications. For this reason, the tube-based approach for MPC and NMPC was proposed in \cite{Mayne2005,Maynenonlineartube} to obtain robust performance of the system. The robust control law is written as follows:
\begin{equation}
u_{i} (t) = \bar{u}_{i} (t) + K_{i} \big(x_{i}(t) - \bar{x}_{i}(t)\big)
\label{robustcontrollaw}
\end{equation}
where $K_{i} \in \mathbb{R}^{m \times n}$ is the feedback gain, $\bar{u}_{i} (t)$ is the output of the DeNMPC, $u_{i} (t)$ is the overall control action applied to the real system, $x_{i}(t) - \bar{x}_{i}(t)$ is the modeling error between the real system and the nominal model for each subsystem.

The uncertainty term for each subsystem which is the summation of the subsystem interaction and the disturbance is written as follows:
\begin{equation}
z_{i}= g_{i} \big(x(t),u(t)\big) + d_{i} (t), \;\;\; i \in \mathbb{I}_{1:N}
\label{z}
\end{equation}
where $z_{i} \in \mathbb{Z}_{i}$ is a robust positively invariant set. It is assumed that $\mathbb{Z}_{i} \subset \mathbb{X}_{i}$ and $K_{i} \mathbb{Z}_{i} \subset \mathbb{U}_{i}$. The nominal state and input have to satisfy:
\begin{eqnarray}
\bar{x}_i & \in & \mathbb{\bar{X}}_{i} = \mathbb{X}_{i} \ominus \mathbb{Z}_{i} \nonumber \\
\bar{u}_i & \in & \mathbb{\bar{U}}_{i} = \mathbb{U}_{i} \ominus K_{i} \mathbb{Z}_{i}
 \end{eqnarray}
where they are in the neighborhoods of the origin.

\begin{figure}[h!]
\centering
  \includegraphics[width=3.2in]{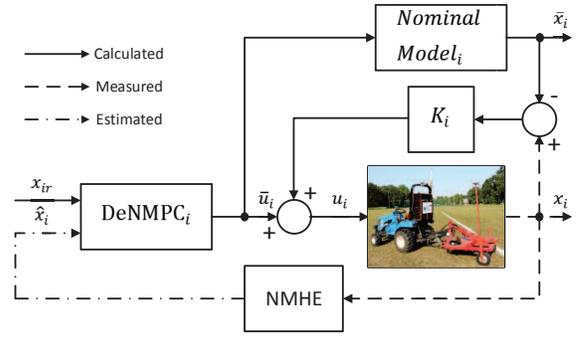}\\
  \caption{The control scheme for ith subsystem}
  \label{controlscheme}
\end{figure}

The nominal controller $\bar{u}_{i} (t)$ is calculated online. However, the ancillary control law $K_{i}$ obtained offline keeps the trajectories of the system error on the robust control invariant set $z_{i}$ centered along the nominal trajectory \cite{Mayne2005}. The control scheme of the system is illustrated in Fig. \ref{controlscheme}.

\subsection{Solution Methods}

The optimization problems in NMHE \eqref{mhe} and in DeNMPC \eqref{DeNMPC} are similar to each other, which makes that the same solution method can be applied for both NMPC and NMHE \cite{TomKraus}. In this paper, the multiple shooting method has been used in a fusion with a generalized Gauss-Newton method. Although the number of iterations cannot be determined in advance, a simple solution was proposed in \cite{Diehl2} in which the number of Gauss-Newton iterations is limited to $1$. Meanwhile, each optimization problem is initialized with the output of the previous one. When implementing the NMPC-NMHE framework for the trajectory tracking problem, the discrete-time optimization is preferred since the trajectory is generally described and stored in discrete time in a spaced based trajectory.

The \emph{ACADO} code generation tool, an open source software package for solving optimization problems \cite{ACADO}, has been used to solve the constrained nonlinear optimization problems in the NMHE and DeNMPC. First, this software generates C-code, which is then converted into a .dll file to be used in \emph{LabVIEW}. Detailed information about the \emph{ACADO} code generation tool can be found in \cite{home,Houska2011a,ACADO}.

%%%%%%%%%%%%%%%%%%%%%%%%%%%%%%%%%%%%%%%%%%%%%%%%%%%%%%%%%%%%%%%%%%%%%%%%%%%%%%%%%%%%%%%%%%%%%%%%%%%%%%%%%%%%%%%%%%%%%%%%%%%%%%%%%%%%%%%%%%%%%%%%%%%%%%%%%%%%%%%%
\section{Experimental Results} \label{realtime}

\subsection{Implementation of NMHE}
Some states of the autonomous tractor-trailer system cannot be measured. Even if states can be measured directly, the obtained measurements contain  time delays and are contaminated with noise. Moreover, data loss from the GPS for global localization of the tractor sometimes occurs. In order to estimate the unmeasurable states or parameters, the NMHE method is used. Since only one GPS antenna is mounted on the tractor and one GPS antenna on the trailer, the yaw angles of the tractor and the trailer cannot be measured. As knowledge of the yaw angles of tractor and trailer is essential for accurate trajectory tracking, these variables have to be accurately estimated.

The inputs to the NMHE algorithm are the position of the tractor, the longitudinal velocity values from the encoders mounted on the rear wheels of the tractor and the steering angle values from the potentiometer mounted on the steering axle of the front wheels of the tractor, the position of the trailer and the steering angle values from the inductive sensors on the trailer and the hitch point angle from the potentiometer between the tractor and the drawbar. The outputs of NMHE are the positions of the tractor and the trailer in the x- and y-coordinate system, the yaw angles for both the tractor and the trailer, the slip coefficients, the hitch point angle and the longitudinal speed. In all the real-time experiments, the estimated values are fed to the robust tube-based DeNMPC .

The NMHE problem is solved at each sampling time with the following constraints on the parameters:
\begin{eqnarray}\label{constraints2}
0.25  \leq & \mu & \leq 1 \nonumber \\
0.25  \leq & \eta & \leq 1 \nonumber \\
0.25  \leq & \kappa & \leq 1
\end{eqnarray}
Even on an ice road, the slip parameters are expected to be around $0.2$. Thus, the lower limit above is chosen for an agricultural operation.

The standard  deviations of the measurements are set to $\sigma_{x^t} = \sigma_{y^t} = \sigma_{x^i} = \sigma_{y^i} = 0.03$ m, $\sigma_{\beta} = 0.0175$ rad, $\sigma_{v} = 0.1$ m/s, $\sigma_{\delta^{t}} = 0.0175$ rad and $\sigma_{\delta^{i}} = 0.0175$ rad based on the information obtained from the real-time experiments. Thus, the following weighting matrix $H$ and the necessary weighting matrix $D_{update}$ to calculate the weighting matrix $P$ have been used in NMHE:
\begin{eqnarray}\label{weightingmatricesVyVu}
H & = & diag(\sigma_{x^t},\sigma_{y^t},\sigma_{x^i},\sigma_{y^i},\sigma_{\beta},\sigma_{v},\sigma_{\delta^{t}},\sigma_{\delta^{i}})^{-1} \nonumber \\
  & = & diag(3,3,3,3,1.75,10,1.75,1.75)^{-1} \times 10^{2}
  \end{eqnarray}
  \begin{eqnarray}
D_{update} & = & diag(x^t, y^t, \theta, x^i, y^i, \psi, \mu, \kappa, \eta, \beta, v) \nonumber \\
      & = & diag(10.0, 10.0, 0.1, 10.0, 10.0, 0.1, \nonumber \\
      && 0.25, 0.25, 0.25, 0.1745, 0.1)
\end{eqnarray}

\subsection{Implementation of Robust Tube-based DeNMPC}

The functions $f_{i}$ and $g_{i}$ in \eqref{nonlinearmodel} are respectively written for the tractor and trailer as follows (subscript 1 refers to the tractor and subscript 2 refers to the trailer):
%\begin{eqnarray}\label{f1g1}
%f_{1}
% =
%\left[
%  \begin{array}{c}
%  \mu v \cos{(\theta)} \\
% \mu v \sin{(\theta)}  \\
%\frac{\mu v \tan{ (\kappa \delta^{t}) } }{L^t} \\
%  \end{array}
%  \right]
%\;\; , \;\;
%g_{1}
% =
%\left[
%  \begin{array}{c}
% 0 \\
% 0 \\
% 0 \\
%  \end{array}
%  \right]
%  \nonumber
%\end{eqnarray}
\begin{eqnarray}\label{f2g2}
f_{1}
 =
\left[
  \begin{array}{c}
  \mu v \cos{(\theta)} \\
 \mu v \sin{(\theta)}  \\
\frac{\mu v \tan{ (\kappa \delta^{t}) } }{L^t} \\
  \end{array}
  \right] \;\; , \;\;
  f_{2}
 =
\left[
  \begin{array}{c}
  \mu v \cos{(\psi)} \\
 \mu v \sin{(\psi)}  \\
\frac{\mu v }{L^i}\big(\sin{(\eta \delta^{i} + \beta)} \\
  \end{array}
  \right] \nonumber \\
g_{1}
 =
\left[
  \begin{array}{c}
 0 \\
 0 \\
 0 \\
  \end{array}
  \right] \mbox{ and } g_{2}
 =
\left[
  \begin{array}{c}
 0 \\
 0 \\
- \frac{l }{L^t} \tan{ (\kappa \delta^{t})} \cos{(\eta \delta^{i} + \beta)} \big) \\
  \end{array}
  \right]
\end{eqnarray}

The DeNMPC problems for the two subsystem are solved at each sampling time with the following constraints on the inputs which are the steering angles of the tractor and the trailer:
\begin{eqnarray}\label{constraints}
-30 \degree  \leq & \delta^{t}(t) & \leq 30 \degree \nonumber \\
-20 \degree  \leq & \delta^{i}(t) & \leq 20 \degree
\end{eqnarray}

The references for the positions and the inputs of the tractor and trailer are respectively changed online while all other references are set to zero as follows:
\begin{eqnarray}\label{}
x_{1r} & = & (x^t_{r},y^t_{r}, \theta_{r})^T \nonumber \\
u_{1r} & = & (\delta^{t}_{ref}) \nonumber \\
x_{2r} & = & (x^i_{r},y^i_{r}, \psi_{r})^T \nonumber \\
u_{2r} & = & (\delta^{i}_{ref})
\end{eqnarray}

The input references are the recent measured (real past) the steering angle of the front wheel of the tractor and the steering angle of the trailer. They are used in the objective function to provide a possibility to penalize the variation in the inputs from time-step to time-step. Moreover, the weighting matrices $Q_i$, $R_i$ and $S_i$ are defined as follows:
\begin{eqnarray}\label{weightingmatricesQRS}
Q_i & = & diag(1,1,0) \nonumber \\
R_i & = & 10 \nonumber \\
S_i & = & diag(10,10,0)
\end{eqnarray}

As can be seen from \eqref{weightingmatricesQRS}, the weighting for the inputs has been chosen large enough in order to get well damped closed-loop behaviour. The reason for such a selection is that since the tractor-trailer system is slow, it cannot give a fast response. Moreover, the weighting values in $S_i$ are set $10$ times larger than the values in the weighting matrix $Q_i$. Thus, the deviations of the predicted values at the end of the horizon from their reference are penalized $10$ times more in the DeNMPC cost function than the previous points.

To handle the uncertainties between the nominal plant and the real-time system for each subsystem, the ancillary control law $K_{i}$ is set to
\begin{eqnarray}\label{ancillarycontrollaw}
K_i  = -diag(\begin{array}{ccc}
         0 & 0 & 3
       \end{array})^{T}
\end{eqnarray}
As can be seen from \eqref{ancillarycontrollaw}, since only the difference between the yaw angles of the nominal model and the real-time system are taken into account, the ancillary control law is linear time invariant. If the differences of x and y-axis would be considered, the ancillary control law should be nonlinear or linear time variant.

%%%%%%%%%%%%%%%%%%%%%%%%%%%%%%%%%%%%%%%%%%%%%%%%%%%%%%%%%%%%%%%%%%%%%%%%%%%%%%%%%%%%%%%%%%%%%%%%%%%%%%%%%%%%%%%%%%%%%%%%%%%%%%%%%%%%%%%%%%%%%%%%%%%%%%%%%%%%%%%%
\subsection{Real-time Results}

A space-based trajectory consisting of three 8-shaped trajectory has been used as a reference signal. Each 8-shaped trajectories consists of two straight lines and two smooth curves. Since the radii of the curves are equal to $10$ m, $8$ m and $6.67$ m, the curvatures of the smooth curves are equal to $0.1$, $0.125$ and $0.15$, respectively. (The curvature of a circle is the inverse of its radius).

The reference generation method in this paper is as follows: As soon as the tractor starts off-track, first, it quickly calculates the closest point on the space-based trajectory. Then, it determines the desired point at a fixed forward distance from the closest point on the trajectory at every specific time instant. While the selection of a large distance from the closest point on the trajectory results in a steady-state error on the trajectory to follow, the drawback of selecting a small distance is that it results in oscillatory behavior of the steering mechanism. Another parameter to determine the mentioned fix forward distance is the longitudinal velocity of the vehicle, \emph{i.e.} the larger longitudinal velocity the larger forward distance. The main goal of the reference generation algorithm for the tractor is both to prevent the oscillations of the steering mechanism and to minimize the steady-state trajectory following error. In this study, this look ahead distance was optimized through trial-and-error and set to $1.6$ m for a forward speed of 1 $m/s$.

\begin{figure}[h!]
\centering
\includegraphics[width=3.2in]{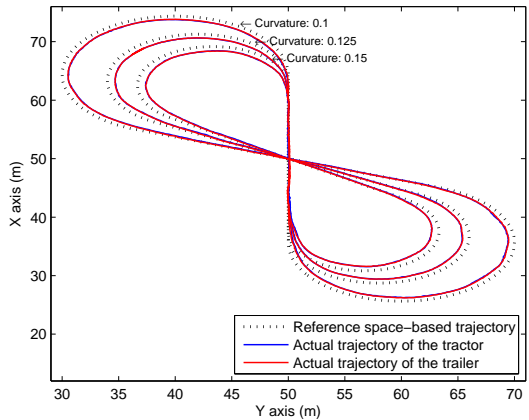}\\
\caption{Reference and actual trajectories}
\label{traj}
\end{figure}

As can be seen from Fig. \ref{traj}, the autonomous tractor-trailer system is capable to stay on-track. In theory, since the reference generation algorithm places the target point $1.6$ meters ahead from the front axle of the tractor, there will be always a steady-state error for the curvilinear trajectories, which makes the tractor "cut corners". On the other hand, no steady-state error is expected for the linear trajectories.

\begin{figure}[t!]
\centering
\includegraphics[width=3.2in]{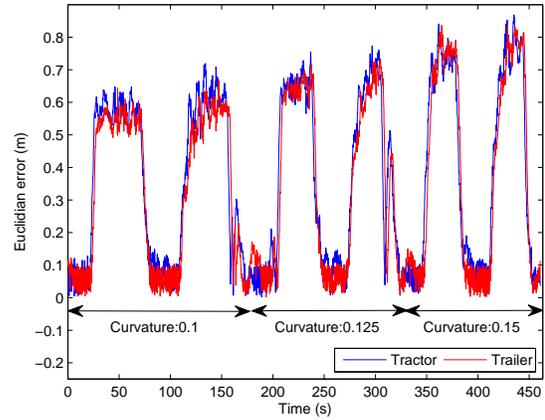}\\
\caption{Euclidian error to the space-based reference trajectory}
\label{error}
\end{figure}

In Fig. \ref{error}, the Euclidian error to the space-based reference trajectory for both the tractor and the trailer is shown. The mean values of the Euclidian error of the tractor and the trailer for the straight lines are $7.95$ cm and $5.42$ cm, respectively. Besides, the mean values of the Euclidian error of the tractor and the trailer for the curvature values $0.1$, $0.125$ and $0.15$ of the curved lines are $59.54$ cm and $55.51$ cm, $66.93$ cm and $64.41$ cm, $76.86$ cm and $76.38$ cm, respectively. Although the robust tube-based DeNMPC  for the trailer calculates the proper outputs for $\delta^{i}$ at $RJ^2$, the error correction for the trailer is limited due to the fact that the length of the drawbar between the tractor and the trailer is only $20$ cm, which corresponded to a maximal lateral displacement of the trailer with respect to the tractor of $10.5$ cm. Moreover, the error correction for the trailer decreases when the curvature value of the curved lines increases. As can be seen from Fig. \ref{error}, if the curvature value of curved lines is equal to or larger than $0.15$, there is no error correction for the trailer due to the mechanical properties of the real-time system.

\begin{figure}[t!]
\centering
\includegraphics[width=3.2in]{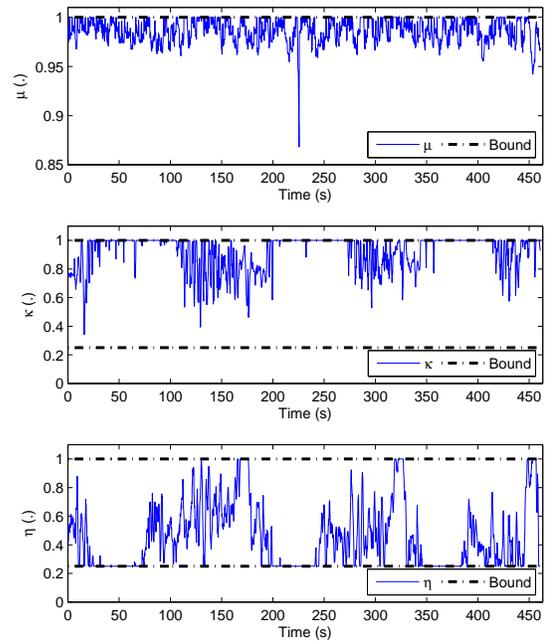}\\
\caption{Tractor longitudinal slip coefficient $(\mu)$, tractor ($\kappa$) and trailer ($\eta$) side slip coefficients}
\label{slips}
\end{figure}

The NMHE parameter estimation performance for the slip coefficients is represented in Fig. \ref{slips}. As can be seen from this figure, the estimated parameter values are within the constraints specified in \eqref{constraints2}. Deviations in the forward slip parameter occurs when a vehicle accelerates, decelerates or soil conditions change, etc. Moreover, the deviations on the side-slip parameters occur each time the steering angles are changed. However, this is not the case in our system. Instead, the deviations in the slip parameters are momentous due to modeling errors in our case.

\begin{figure}[t!]
\centering
\includegraphics[width=3.2in]{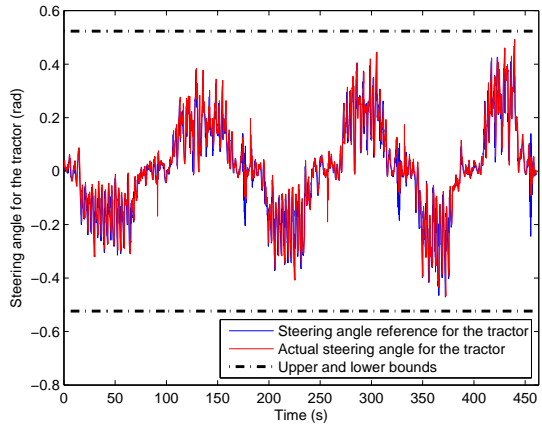}\\
\caption{Reference and actual steering angle for the tractor}
\label{steering_tra}
\end{figure}

\begin{figure}[t!]
\centering
\includegraphics[width=3.2in]{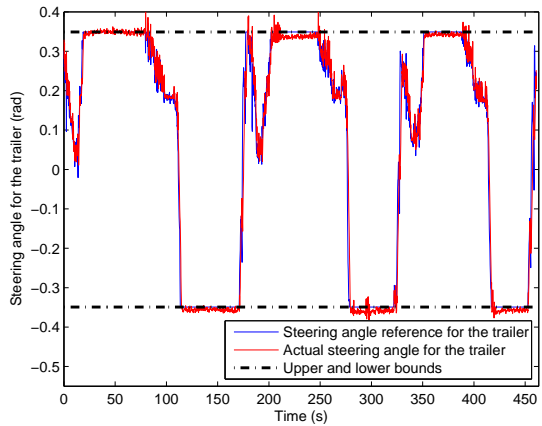}\\
\caption{Reference and actual steering angle for the trailer}
\label{steering_imp}
\end{figure}

In Figs. \ref{steering_tra}-\ref{steering_imp}, the outputs, the steering angle ($\delta^{t}$) reference for the tractor and the steering angle ($\delta^{i}$) reference for the trailer, of the robust tube-based DeNMPC  are illustrated. As can be seen from these figures, the performance of the low level controllers is sufficient. Moreover, it is observed from Fig. \ref{steering_imp} that even if the output of the robust tube-based DeNMPC  for the trailer reaches its constraints, the error correction is limited due to the aforementioned limited length of the drawbar. It is to be noted that while the contribution of the state-feedback controller is less than $1\%$ to the overall control signal for the tractor, it is around $5\%$ for the trailer since the influence of the tractor steering angle on the yaw angle of the trailer is neglected, as explained in Section \ref{sectionDeNMPC}.


The execution times for DeNMPC for the tractor and trailer and centralized NMPC (CeNMPC) are given in Table \ref{mpcmheperformance}. During the real-time experiments, a real-time controller equipped with a 2.26 GHz Intel Core 2 Quad Q9100 quad-core processor (NI PXI-8110, National Instruments, Austin, TX, USA) has been used. The NMHE and NMPC routine was assigned to one core. As can be seen from this table, in DeNMPC the computation time needed to solve the optimization problem was always below 1.5 ms for both the NMPC for the tractor and the trailer. When these are summed, the overall computation time is less than half the time needed for the CeNMPC. However, it should be noted here that the maximum computation time of 7.24 needed for the CeNMPC would still be acceptable in this application \cite{erkanCeNMPC}. Since the optimization problem in DeNMPCs is relatively simpler than the one in CeNMPC, the computation time of DeNMPCs is less than the one in CeNMPC as well.

\begin{table}
\centering
\caption{Execution times of DeNMPCs for the tractor and trailer and CeNMPC}\label{mpcmheperformance}
\begin{tabular}{lccc}
  \hline
   &  Minimum & Average &  Maximum (ms)\\
   &  (ms) &   (ms) &  (ms) \\
       \hline
  DeNMPC for the tractor & &  &  \\
  Preparation & 1.1791 & 1.1816 & 1.3199 \\
  Feedback & 0.0293 &  0.0313 &  0.1173 \\
  Overall & 1.2084 &  1.2129 &  1.4372 \\
  \hline
      \hline
  DeNMPC for the trailer & &  &  \\
  Preparation & 1.2487 & 1.2540 & 1.3234 \\
  Feedback & 0.0288 &  0.0541 &  0.1505 \\
  Overall & 1.2775 &  1.3081 &  1.4739 \\
  \hline
    \hline
  CeNMPC  & &  &  \\
  Preparation & 6.5462 & 6.6632 & 6.9260 \\
  Feedback & 0.0521 &  0.1345 &  0.3140 \\
  Overall & 6.5983 &  6.7977 &  7.2400 \\
  \hline
\end{tabular}
\end{table}

%%%%%%%%%%%%%%%%%%%%%%%%%%%%%%%%%%%%%%%%%%%%%%%%%%%%%%%%%%%%%%%%%%%%%%%%%%%%%%%%%%%%%%%%%%%%%%%%%%%%%%%%%%%%%%%%%%%%%%%%%%%%%%%%%%%%%%%%%%%%%%%%%%%%%%%%%%%%%%%%

\section{Conclusions and Future Research} \label{Conc}

In this study, a fast robust tube-based DeNMPC-NMHE framework based-on an adaptive tricycle kinematic model has been elaborated for the control of an autonomous tractor-trailer system. The experimental results in the field have shown that the NMHE is able to accurately estimate the unmeasurable states and parameters online, and the robust tube-based DeNMPC  is robust against neglecting subsystem interactions and uncertainties. The mean value of the Euclidian error to the straight line was $7.95$ cm and $5.42$ cm for the tractor and trailer, respectively. It is to be noted that the \emph{ACADO} code generation provide feedback in the millisecond range for DeNMPC so that the DeNMPC needed less than 75$\%$ of the the computation time required for CeNMPC.

\subsection{Future research}

Since the robust tube-based DeNMPC-NMHE framework based upon the adaptive kinematic model of the tractor-trailer system provides feedback times in a millisecond, it is amenable to extend this framework based-on a dynamic model.

%
%\appendices
%\section{Proof of the First Zonklar Equation}
%Appendix one text goes here.
%\section{}
%Appendix two text goes here.

%%%%%%%%%%%%%%%%%%%%%%%%%%%%%%%%%%%%%%%%%%%%%%%%%%%%%%%%%%%%%%%%%%%%%%%%%%%%%%%%%%%%%%%%%%%%%%%%%%%%%%%%%%%%%%%%%%%%%%%%%%%%%%%%%%%%%%%%%%%%%%%%%%%%%%%%%%%%%%%%

\section*{Acknowledgment}
This work has been carried out within the IWT-SBO 80032 (LeCoPro) project funded by the Institute for the Promotion of Innovation through Science and Technology in Flanders (IWT-Vlaanderen).

\ifCLASSOPTIONcaptionsoff
  \newpage
\fi

\bibliography{references_file}
\bibliographystyle{IEEEtran}

\begin{IEEEbiography}[{\includegraphics[width=1in,height=1.25in,clip,keepaspectratio]{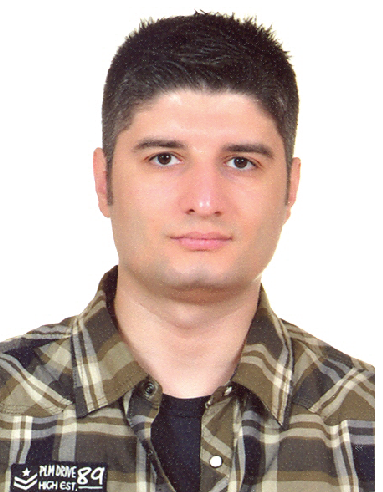}}]{Erdal Kayacan} (S\textquoteright 06-SM\textquoteright 12)  was born in Istanbul, Turkey on January 7, 1980. He received a B.Sc. degree in electrical engineering from in 2003 from Istanbul Technical University in Istanbul, Turkey as well as a M.Sc. degree in systems and control engineering in 2006 from Bogazici University in Istanbul, Turkey. In September 2011, he received a Ph.D. degree in electrical and electronic engineering at Bogazici University in Istanbul, Turkey. After finishing his post-doctoral research in KU Leuven at the division of mechatronics, biostatistics and sensors (MeBioS), he is currently pursuing his research in Nanyang Technological University at the School of Mechanical and Aerospace Engineering as an assistant professor. His research areas are unmanned aerial vehicles, robotics, mechatronics, soft computing methods, sliding mode control and model predictive control.

Dr. Kayacan has been serving as an editor in Journal on Automation and Control Engineering (JACE) and editorial advisory board in Grey Systems Theory and Application.
\end{IEEEbiography}

\begin{IEEEbiography}[{\includegraphics[width=1in,height=1.25in,clip,keepaspectratio]{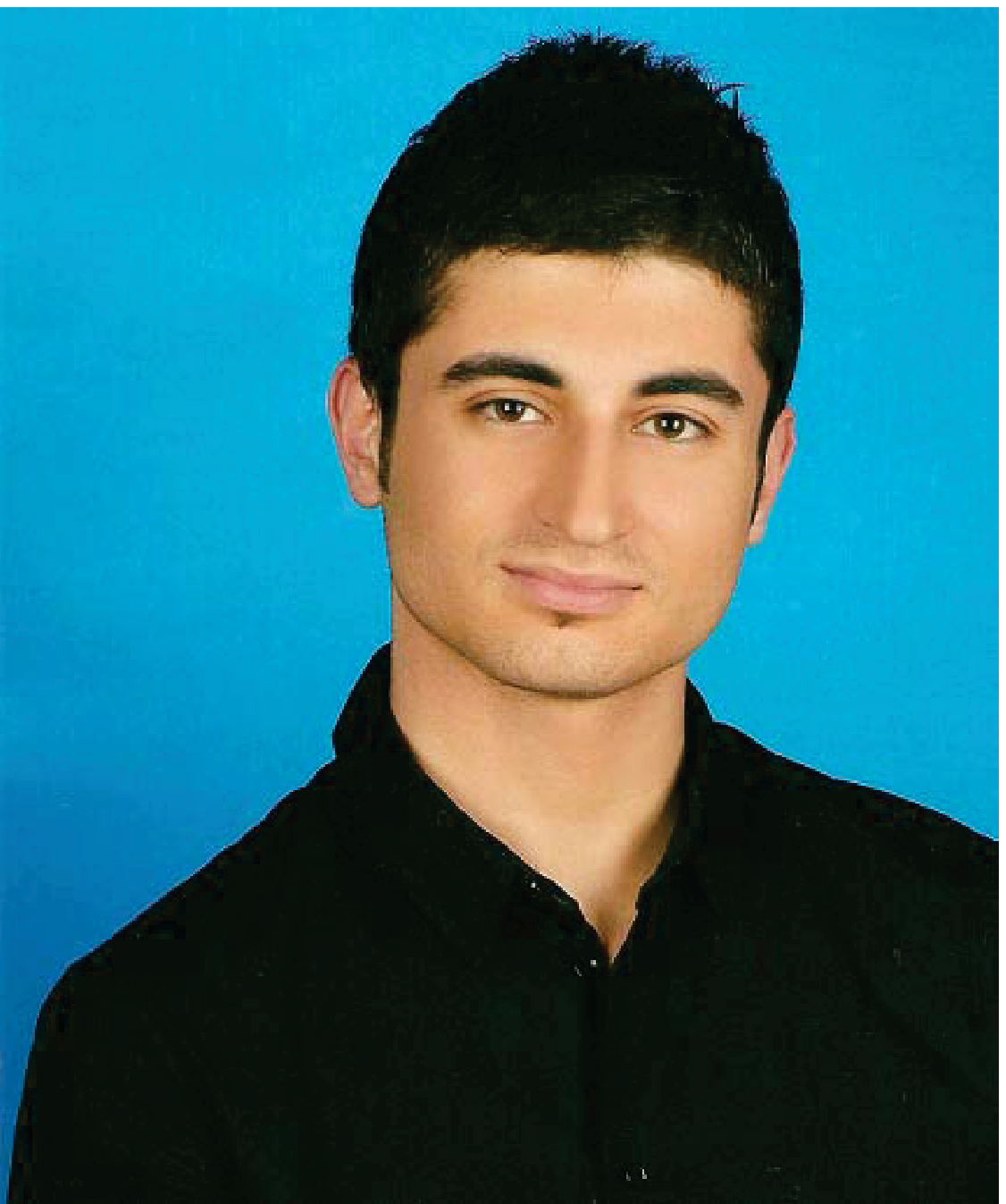}}]{Erkan Kayacan} (S\textquoteright 12) was born in Istanbul, Turkey, on April 17, 1985. He received the B.Sc. and the M.Sc. degrees in mechanical engineering from Istanbul Technical University, Istanbul, in 2008 and 2010, respectively. He is a PhD student and research assistant at University of Leuven (KU Leuven) in the division of mechatronics, biostatistics and sensors (MeBioS). His research interests include model predictive control, moving horizon estimation, distributed and decentralized control, intelligent control, vehicle dynamics and mechatronics.
\end{IEEEbiography}

\begin{IEEEbiography}[{\includegraphics[width=1in,height=1.25in,clip,keepaspectratio]{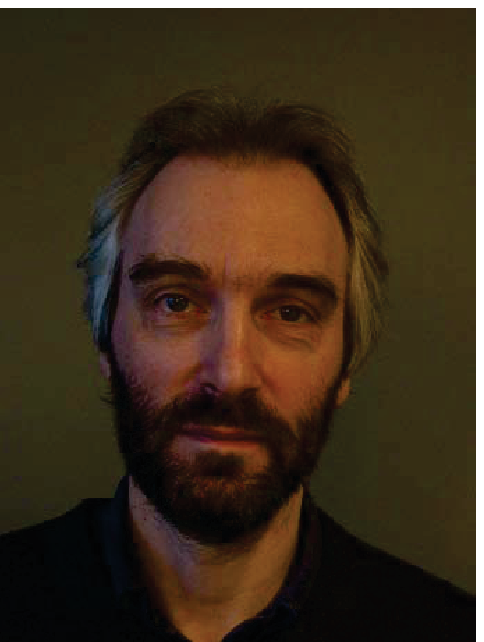}}]{Herman Ramon} graduated as an agricultural engineer from Gent University. In 1993 he obtained a Ph.D. in applied biological sciences at the Katholieke Universiteit Leuven. He is currently Professor at the Faculty of Agricultural and Applied Biological Sciences of the Katholieke Universiteit Leuven, lecturing on agricultural machinery and mechatronic systems for agricultural machinery. He has a strong research interest in precision technologies and advanced mechatronic systems for processes involved in the production chain of food and nonfood materials, from the field to the end user. He is author or co-author of more than 90 papers.
\end{IEEEbiography}

\begin{IEEEbiography}[{\includegraphics[width=1in,height=1.25in,clip,keepaspectratio]{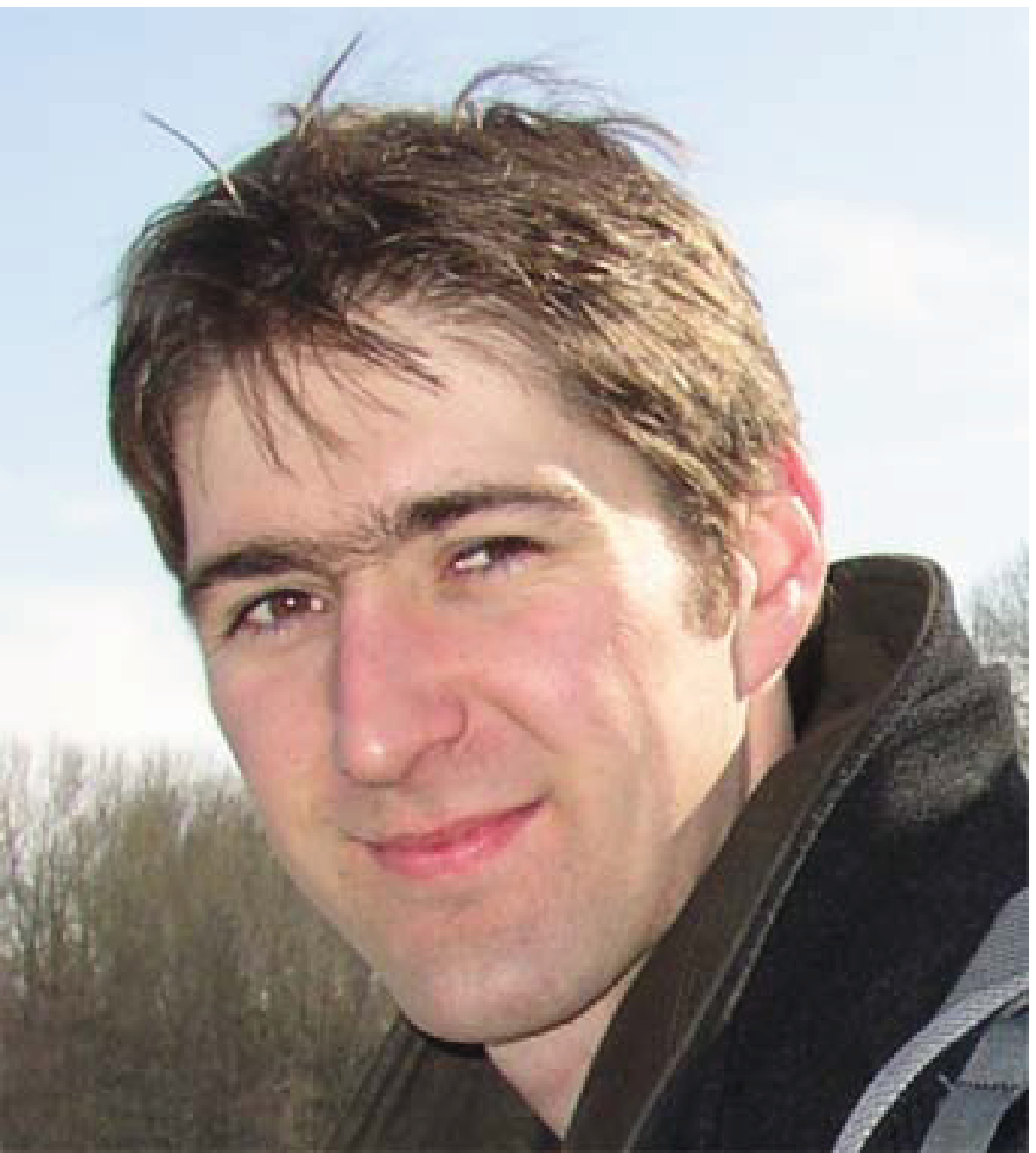}}]{Wouter Saeys} is currently Assistant Professor in Biosystems Engineering at the Department of Biosystems of the University of Leuven in Belgium. He obtained his Ph.D. at the same institute and was a visiting postdoc at the School for Chemical Engineering and Advanced Materials of the University of Newcastle upon Tyne, UK and at the Norwegian Food Research Institute - Nofima Mat in Norway. His main research interests are optical sensing, process monitoring and control with applications in food and agriculture. He is author of 50 articles (ISI) and member of the editorial board of Biosystems Engineering.
\end{IEEEbiography}

\end{document}